\documentclass[10pt,a4paper]{book}
\usepackage{udineHyderabad}
\begin{document}
\talktitle{Cosmological  Theories of Special and General Relativity - II}

\talkauthors{Moshe Carmeli \structure{a}}
\authorstucture[a]{Department of Physics, 
                   Ben Gurion University  of the Negev, 
                   Beer Sheva 84105, Israel}

\shorttitle{Theories of Special and General Relativity - II} 

\firstauthor{M. Carmeli}

\begin{abstract}
Astronomers measure distances to faraway galaxies and their velocities. They 
do that in order to determine the expansion rate of the Universe. In Part I of 
these lectures the foundations of the theory of the expansion of the Universe
was given. In this part we present the theory. A formula for the distance
of the galaxy in terms of its velocity is given. It is very simple: 
$r(v)=c\tau/\beta\sinh\beta v/c$, where $\tau$ is the Big Bang time, 
$\beta=\sqrt{1-\Omega_m}$, and $\Omega_m$ is the mass density of the Universe. 
For $\Omega_m<1$ this 
formula clearly indicates that the Universe is expanding with acceleration, 
as experiments clearly show.
\end{abstract}

\section{Gravitational field equations}
In the four-dimensional spacevelocity the spherically symmetric metric is 
given by
$$ds^2=\tau^2dv^2-e^\mu dr^2-R^2\left(d\theta^2+\sin^2\theta d\phi^2\right),
\eqno(1.1)$$
where $\mu$ and $R$ are functions of $v$ and $r$ alone, and comoving 
coordinates $x^\mu=(x^0,x^1,x^2,x^3)=(\tau v,r,\theta,\phi)$ have been used. 
With the above choice of coordinates, the zero-component of the geodesic
equation becomes an identity, and since $r$, $\theta$ and $\phi$ are constants
along the geodesics, one has $dx^0=ds$ and therefore
$u^\alpha=u_\alpha=\left(1,0,0,0\right).$
The metric (1.1) shows that the area of the sphere $r=constant$ 
is given by
$4\pi R^2$ and that $R$ should satisfy $R'=\partial R/\partial r>0$. The
possibility that $R'=0$ at a point $r_0$ is excluded since it would
allow the lines $r=constants$ at the neighboring points $r_0$ and $r_0+dr$ to
coincide at $r_0$, thus creating a caustic surface at which the comoving 
coordinates break down.

As has been shown in Part I the Universe expands by the null
condition $ds=0$, and if the expansion is spherically symmetric one has
$d\theta=d\phi=0$. The metric (1.1) then yields
$\tau^2 dv^2-e^\mu dr^2=0,$
thus
$$\frac{dr}{dv}=\tau e^{-\mu/2}.\eqno(1.2)$$
This is the differential equation that determines the Universe expansion. In
the following we solve the gravitational field equations in order to find out
the function $\mu\left(r.v\right)$.

The gravitational field equations, written in the form
$$R_{\mu\nu}=\kappa\left(T_{\mu\nu}-g_{\mu\nu}T/2\right),\eqno(1.3)$$
where 
$$T_{\mu\nu}=\rho_{eff}u_\mu u_\nu+p\left(u_\mu u_\nu-g_{\mu\nu}\right),
\eqno(1.4)$$ 
with $\rho_{eff}=\rho-\rho_c$ and $T=T_{\mu\nu}g^{\mu\nu}$, are now solved.
One finds that the only nonvanishing components of $T_{\mu\nu}$ 
are $T_{00}=\tau^2\rho_{eff}$, $T_{11}=c^{-1}\tau pe^\mu$, $T_{22}=c^{-1}\tau 
pR^2$ and $T_{33}=c^{-1}\tau pR^2\sin^2\theta$, and that $T=\tau^2\rho_{eff}-
3c^{-1}\tau p$.

One obtains three independent field equations (dot and prime denote derivatives
with $v$ and $r$)  
$$e^\mu\left(2R\ddot{R}+\dot{R}^2+1\right)-R'^2=-\kappa\tau c^{-1} e^\mu R^2p,
\eqno(1.5)$$
$$2\dot{R}'-R'\dot{\mu}=0,\eqno(1.6)$$ 
$$e^{-\mu}\left[\frac{1}{R}R'\mu'-\left(\frac{R'}{R}\right)^2-\frac{2}{R}R''
\right]+\frac{1}{R}\dot{R}\dot{\mu}+\left(\frac{\dot{R}}{R}\right)^2+
\frac{1}{R^2}=\kappa\tau^2\rho_{eff}.\eqno(1.7)$$ 
\section{Solution of the field equations}
The solution of (1.6) satisfying the condition $R'>0$ is given by
$$e^\mu=R'^2/\left(1+f\left(r\right)\right),\eqno(2.1)$$
where $f\left(r\right)$ is an arbitrary function of the coordinate $r$ and 
satisfies the
condition $f\left(r\right)+1>0$. Substituting (2.1) in the other two 
field equations (1.5) and (1.7) then gives
$$2R\ddot{R}+\dot{R}^2-f=-\kappa c^{-1}\tau R^2p,\eqno(2.2)$$ 
$$\frac{1}{RR'}\left(2\dot{R}\dot{R'}-f'\right)+\frac{1}{R^2}\left(\dot{R}^2-f
\right)=\kappa\tau^2\rho_{eff},\eqno(2.3)$$
respectively.

The simplest solution of the above two equations, which satisfies the 
condition $R'=1>0$, is given by $R=r$.
Using this in Eqs. (2.2) and (2.3) gives
$f\left(r\right)=\kappa c^{-1}\tau pr^2$, and
$f'+f/r=-\kappa\tau^2\rho_{eff}r$,
respectively. 
Using the values of $\kappa=8\pi G/c^2\tau^2$ and $\rho_c=3/8\pi G\tau^2$, we
obtain
$$f\left(r\right)=\left(1-\Omega_m\right)r^2/c^2\tau^2,\eqno(2.4)$$
where $\Omega_m=\rho/\rho_c$. We also obtain
$$p=\frac{1-\Omega_m}{\kappa c\tau^3}=\frac{c}{\tau}\frac{1-\Omega_m}{8\pi G}
=4.544\left(1-\Omega_m\right)\times 10^{-2} g/cm^2,\eqno(2.5)$$
$$e^{-\mu}=1+f\left(r\right)=1+\tau c^{-1}\kappa pr^2=1+
\left(1-\Omega_m\right)r^2/c^2\tau^2.\eqno(2.6)$$

Accordingly, the line element of the Universe is given by
$$ds^2=\tau^2dv^2-\frac{dr^2}{1+\left(1-\Omega\right)r^2/c^2\tau^2}
-r^2\left(d\theta^2+\sin^2\theta d\phi^2\right),\eqno(2.7)$$
or,
$$ds^2=\tau^2dv^2-\frac{dr^2}{1+\left(\kappa\tau/c\right)pr^2}
-r^2\left(d\theta^2+\sin^2\theta d\phi^2\right).\eqno(2.8)$$
This line element is the comparable to the FRW line element in the standard theory.

It  will be recalled that the Universe expansion is determined by Eq. (1.2),
$dr/dv=\tau e^{-\mu/2}$. The only thing that is left to be determined is the
sign of $(1-\Omega_m)$ or the pressure $p$. Thus we have
$$\frac{dr}{dv}=\tau\sqrt{1+\kappa\tau c^{-1}pr^2}=\tau\sqrt{1+
\frac{1-\Omega_m}{c^2\tau^2}r^2}.\eqno(2.9)$$
\section{Physical meaning}
For $\Omega_m>1$ one obtains
$$r\left(v\right)=\frac{c\tau}{\alpha}\sin\alpha\frac{v}{c},\hspace{5mm}
\alpha=\sqrt{\Omega_m-1}.\eqno(3.1)$$
This is obviously a closed Universe, and presents a decelerating expansion.

For $\Omega_m<1$ one obtains
$$r\left(v\right)=\frac{c\tau}{\beta}\sinh\beta\frac{v}{c},\hspace{5mm}
\beta=\sqrt{1-\Omega_m}.\eqno(3.2)$$
This is now an open accelerating Universe.

For $\Omega_m=1$ we have, of course, $r=\tau v$.
\section{The accelerating Universe}
From the above one can write the expansion of the Universe in the standard
Hubble form $v=H_0r$ with 
$$H_0=h\left[1-\left(1-\Omega_m\right)v^2/6c^2\right],\eqno(4.1)$$
where $h=\tau^{-1}$. Thus $H_0$ depends on the distance it is being 
measured \cite{Peebles}.
It is well-known that the farther the distance, the lower the value for $H_0$
is measured. This is possible only for $\Omega_m<1$, i.e. when the Universe 
is accelerating. In that case the pressure is positive. 

Figure 1 describes the Hubble diagram of the above solutions for the three 
types of expansion for values of $\Omega_m$ from 100 to 0.245. The figure
describes the three-phase evolution of the Universe. Curves (1)-(5) represent
the stages of {\it decelerating expansion} according to 
Eq. (3.1). As the density 
of matter $\rho$ decreases, the Universe goes over from the lower curves to 
the upper ones, but it does not have enough time to close up to a Big Crunch.
The Universe subsequently goes over to curve (6) with $\Omega_m=1$, at which time
it has a constant expansion for a fraction of a second. This then followed
by going to the upper curves (7) and (8) with $\Omega_m<1$, where the Universe
expands with {\it acceleration} according to Eq. (3.2). Curve no. 8 fits
the present situation of the Universe. For curves (1)-(4) in the diagram we
use the cutoff when the curves were at their maximum.  

\begin{figure}[t]
\centering
\includegraphics[width=10cm]{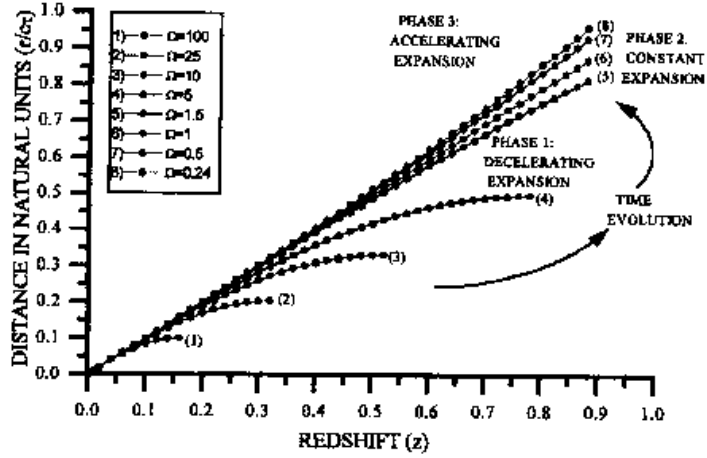}
\caption{Hubble's diagram describing the three-phase evolution of the
Universe according to cosmological general relativity theory.}
\label{fig.1}
\end{figure}
\section{Theory versus experiment}
To find out the numerical value of $\tau$ we use the relationship between
$h=\tau^{-1}$ and $H_0$ given by Eq. (4.1)(CR denote values according to 
Cosmological Relativity):
$$H_0=h\left[1-\left(1-\Omega_m^{CR}\right)z^2/6\right],\eqno(5.1)$$
where $z=v/c$ is the redshift and $\Omega_m^{CR}=\rho_m/\rho_c$ with 
$\rho_c=3h^2/8\pi G$. (Notice that our $\rho_c=1.194\times 10^{-29}g/cm^3$ is different from 
the standard $\rho_c$ defined with $H_0$.) The redshift parameter $z$ 
determines the distance at which $H_0$ is measured. We choose $z=1$ and take 
for $\Omega_m^{CR}=0.245$,
its value at the present time (corresponds to 0.32 in the 
standard theory), Eq. (5.1) then gives $H_0=0.874h.$
At $z=1$ the corresponding Hubble parameter $H_0$ according to the 
latest results from HST can be taken \cite{Freed} as $H_0=70$km/s-Mpc, thus 
$h=(70/0.874)$km/s-Mpc, or $h=80.092\mbox{\rm km/s-Mpc},$
and $\tau=12.486 Gyr=3.938\times 10^{17}s.$

What is left is to find the value of $\Omega_\Lambda^{CR}$. We have 
$\Omega_\Lambda^{CR}=\rho_c^{ST}/\rho_c$, where $\rho_c^{ST}=3H_0^2/8\pi 
G$ and $\rho_c=3h^2/8\pi G$. Thus $\Omega_\Lambda^{CR}=(H_0/h)^2=0.874^2$,
or $\Omega_\Lambda^{CR}=0.764.$
As is seen from the above equations one has 
$$\Omega_T=\Omega_m^{CR}+\Omega_\Lambda^{CR}=0.245+0.764=1.009\approx 1,
\eqno(5.2)$$
which means the Universe is Euclidean.

Our results confirm those of the supernovae experiments and indicate on the
existance of the dark energy as has recently received confirmation from the
Boomerang cosmic microwave background experiment \cite{Bernard,Bond}, which showed that 
the Universe is Euclidean.
\section{Comparison with general relativity}
One has to add the time coordinate and the result is a five-dimensional theory
of space-time-velocity. One can show that all the classical experiments 
predicted by general relativity are also predicted by CGR. Also predicted a
wave equation for gravitational radiation. In the linear approximation one
obtains
$$\left(\frac{1}{c^2}\frac{\partial^2}{\partial t^2}-\nabla^2+\frac{1}{\tau^2}
\frac{\partial^2}{\partial v^2}\right)\gamma_{\mu\nu}=-2\kappa T_{\mu\nu},
\eqno(6.1)$$
where $\gamma_{\mu\nu}$ is a first approximation term,
$$g_{\mu\nu}\approx\eta_{\mu\nu}+h_{\mu\nu}=\eta_{\mu\nu}+\gamma_{\mu\nu}-
\eta_{\mu\nu}\gamma/2,\eqno(6.2)$$
$$\gamma=\eta^{\alpha\beta}\gamma_{\alpha\beta}.\eqno(6.3)$$
Hence CGR predicts that gravitational waves depend not only on space and time
but also on the redshift of the emitting source.
\begin{table}[h]
\begin{center}  
\begin{tabular}{p{35mm}p{35mm}p{35mm}}
\hline\\
&COSMOLOGICAL&STANDARD\\
&RELATIVITY&THEORY\\
\hline\\
Theory type&Spacevelocity&Spacetime\\
Expansion&Tri-phase:&One phase\\
type&decelerating, constant,&\\
&accelerating&\\
Present expansion&Accelerating&One of three\\
&(predicted)&possibilities\\
Pressure&$0.034g/cm^2$&Negative\\
Cosmological constant&$1.934\times 10^{-35}s^{-2}$&Depends\\
&(predicted)&\\
$\Omega_T=\Omega_m+\Omega_\Lambda$&1.009&Depends\\
Constant-expansion&8.5Gyr ago&No prediction\\
occurs at&(Gravity is included)&\\
Constant-expansion&Fraction of&Not known\\
duration&second&\\
Temperature at&146K&No prediction\\
constant expansion&(Gravity is included)&\\
\hline
\end{tabular}
\caption{Cosmological parameters in cosmological general  relativity and in
standard theory.   \label{t2}}
\end{center}
\end{table}
\section{New developments on dark matter} 
Using the theory presented here, John Hartnett has recently shown that there
is no need for the existence of dark matter in spiral galaxies. We only give
the references to this work by Hartnett \cite{Hartnett1,Hartnett2,Hartnett3}.

\end{document}